\documentclass[aps,prd,showpacs,superscriptaddress,nofootinbib]{revtex4-2} 
\usepackage{amsmath,amssymb,bm,graphicx}
\usepackage{xcolor}
\usepackage[hidelinks]{hyperref}
\usepackage{siunitx}
\usepackage{placeins}
\usepackage{tabularray}
\UseTblrLibrary{booktabs}

\usepackage{orcidlink}
\DeclareUnicodeCharacter{2212}{$-$}

\begin{document}

\title{Revisiting the Duff Argument in Cosmology:\\ From Dimensional Quantities to Cross-Epoch Observables}
\author{Seokcheon Lee \orcidlink{0000-0003-0861-1300}}
\email{skylee@skku.edu}
\affiliation{Department of Physics, Institute of Basic Science, Sungkyunkwan University, Suwon 16419, Korea}

\begin{abstract}
The standard Duff-type argument states that the numerical variation of a dimensional constant is unit-dependent when considered in isolation from the physical standards used to compare it, and is therefore not by itself an observable. We ask whether this conclusion is sufficient to determine the observational status of complete relations connecting different cosmological epochs. In a Robertson--Walker spacetime, the cosmological principle and Weyl's postulate fix the homogeneous geometry and fundamental timelike congruence without uniquely fixing the temporal rate along it. The Hamiltonian constraint likewise does not dynamically select that rate. Consequently, Robertson--Walker spacetime admits a family of homogeneous temporal parametrizations. Generalized Cosmological Time retains a specified common member, $\mathcal T_G$, related to proper time by $d\tau=\mathcal N_G\,d\mathcal T_G$, with $(\tau,1)$ as the unit-rate special member of the same class. In the minimally extended varying speed of light framework, length--mass--time quantities obey correlated slice-wise scalings chosen to preserve the same local physical laws on every homogeneous hypersurface. These scalings form a unit-rescaling structure, so purely local dimensionless observables are insensitive to $\mathcal N_G$. Cross-epoch observables require more than such local scalings. They also require local reference standards, a propagation or comparison map, and, when distinct source populations are compared, a source-equivalence prescription. Cosmological time dilation illustrates this distinction: its interpretation depends on source standardization and residual source evolution. We distinguish the unit dependence of individual dimensional quantities from the observational content of complete cross-epoch relations. The former does not by itself determine the latter.
\end{abstract}


\maketitle

\vspace{1em}

\section{Introduction}
\label{sec:introduction}

The possible cosmological evolution of quantities conventionally regarded as fundamental constants has long been discussed in gravitational and cosmological physics. Examples include the speed of light $c$, Newton's constant $G$, Planck's constant $\hbar$, and dimensionless combinations such as the fine-structure constant $\alpha$~\cite{Uzan:2002vq,Magueijo:2003gj}. A central conceptual point, emphasized particularly by Duff and collaborators, is that the numerical value of a dimensional quantity depends on the adopted units. An isolated statement such as ``$c$ has changed'' therefore does not by itself define a unit-independent observable. Measurable statements must instead be formulated through dimensionless quantities or other unit-independent relations~\cite{Duff:2001ba,Duff:2002vp}. We take this conclusion as a premise rather than as a point to be disputed.

Related questions about physical standards, unit conventions, and cosmological time have a longer history. Here, physical standards mean locally realized reference quantities used to assign numerical values to measurements, such as clock transitions, length and mass standards, and calibrated spectral or detector scales. Dirac distinguished atomic and gravitational measures of length and time in his cosmological construction, while Dicke emphasized the invariance of physical laws under coordinate-dependent transformations of units. Canuto \textit{et al.} developed a scale-covariant framework that distinguishes atomic and gravitational systems of units. Ellis and Uzan later emphasized the different operational roles assigned to the symbol $c$ in varying-speed-of-light discussions~\cite{Dirac:1938mt,Dicke:1961gz,Canuto:1977dp,Ellis:2003pw}. The temporal parameter $\mathcal T_G$ introduced here is not identified with either unit system used in the earlier scale-covariant literature. It is instead a scalar temporal parametrization of the same Robertson--Walker (RW) fundamental congruence, with its rate encoded by $\mathcal N_G$, rather than a separate physical system of rods and clocks. The present question is whether the standard unit-dependence conclusion for isolated dimensional quantities is sufficient to determine a complete comparison between standards realized at different cosmological epochs. Such a comparison requires the relevant standards, propagation or comparison map, and any cross-epoch identification rule to be specified together. The purpose is to determine whether correlated temporal dependence survives or cancels in that complete relation, complementing rather than rejecting the Duff argument.

The relevant geometric setting is the temporal reparametrization freedom of RW spacetime. The cosmological principle (CP) and Weyl's postulate determine the homogeneous and isotropic geometry and identify the fundamental timelike congruence, but they do not uniquely fix the temporal rate along it. Standard treatments commonly adopt comoving proper time and the unit-lapse representation, while the same RW spacetime admits a class of smooth monotonic temporal parametrizations with a time-dependent lapse~\cite{Ellis:1998ct,Islam2002,Gourgoulhon:2007ue,Ryder2009,Katanaev:2016byi}. A lapse associated only with a relabeling of the temporal coordinate remains gauge and introduces no independent propagating degree of freedom~\cite{Gourgoulhon:2007ue,Arnowitt:1962hi,Cook:2000vr,Lee:2024zcu}.

Earlier minimally extended varying speed of light (meVSL) discussions emphasized the complementary slice-wise point that spatial constancy on one homogeneous hypersurface does not require dimensional quantities to have identical numerical realizations on different hypersurfaces~\cite{Lee:2023bjz,Lee:2024mal}. The present work sharpens this observation in the temporal sector by separating RW lapse freedom from the comparison of endpoint rates defined with respect to one specified common temporal parameter. Generalized Cosmological Time (GCT) retains a common cosmological time parameter $\mathcal T_G$ within this RW temporal structure~\cite{Lee:2025osx,Lee:2026kyz}. Along the fundamental congruence, $d\tau=\mathcal N_G\,d\mathcal T_G$, with $\mathcal N_G=(U^\mu\nabla_\mu\mathcal T_G)^{-1}$. The pair $(\mathcal T_G,\mathcal N_G)=(\tau,1)$ is the proper-time representation and the unit-rate member of the same temporal-parametrization class. For a specified $\mathcal T_G$, $\mathcal N_G$ is its temporal-rate scalar; in coordinates adapted to $\mathcal T_G$, it is represented by the ordinary RW lapse~\cite{Ellis:1998ct,Gourgoulhon:2007ue,Lee:2026kyz}. Covariance fixes the representation of this relation but does not select a particular function $\mathcal N_G(a)$. RW geometry and the reparametrization-invariant gravitational sector do not dynamically select a function $\mathcal N_G(a)$; a non-unit form must be specified phenomenologically and constrained through the cross-epoch observables to which it contributes.

The meVSL framework supplies one phenomenological realization, $\mathcal N_G(a)=a^{b/4}$, together with correlated cross-epoch scalings of dimensional quantities~\cite{Lee:2024zcu,Lee:2023bjz,Lee:2024mal,Lee:2025osx,Lee:2020zts}. Their correlated scalings are chosen so that the local physical relations considered in the framework retain their standard form on each homogeneous hypersurface. Previous GCT work applied the same temporal-rate relation to supernovae (SNe), gamma-ray bursts (GRBs), and quasars (QSOs) time-dilation signals~\cite{Lee:2026kyz}. We use this setting to organize three distinct observable outcomes: exact cancellation of correlated dimensional factors, conditional survival of an endpoint temporal-rate ratio, and additional model dependence requiring separate microphysical or source modeling. Cosmological time dilation (CTD) provides the explicit timing example; its propagation law remains standard, while inference of a temporal-rate contribution requires a source-equivalence prescription and is degenerate with residual source evolution.

Section~\ref{sec:duff_cross_epoch} isolates the Duff-type argument from the cross-epoch question. Section~\ref{sec:lapse_to_gct} develops the RW/GCT temporal structure. Section~\ref{sec:observables} presents the CTD relation and source-equivalence condition. Section~\ref{sec:phenomenology} summarizes the meVSL realization and the correlated dimensional sector. Detailed covariance and lapse-variation derivations are collected in the Appendices.

\section{From isolated dimensional variation to cross-epoch observability}
\label{sec:duff_cross_epoch} 

The unit-dependence argument and the cross-epoch question should be kept logically separate. Let $Q$ denote a dimensional quantity. A difference in the numerical value of $Q$ between two epochs is not invariant under a change of units and cannot constitute an observable solely through that numerical difference~\cite{Duff:2001ba,Duff:2002vp}.

A cosmological observation, however, does not measure an isolated number $Q_i$ on a hypersurface $\Sigma_i$. It measures a relation among physical standards, propagation effects, source properties, detector standards, and the temporal or spatial mapping between emission and observation. Unit-dependent factors may cancel in such a relation, leaving a dimensionless or otherwise operationally invariant quantity. They may also cancel completely, leaving no sensitivity to the underlying parametrization. Which outcome occurs is a property of the full observable and cannot be inferred from the isolated scaling of one dimensional quantity.

A simple example is provided by photon propagation. The wavelength and proper-time frequency are both dimensionful quantities and evolve cosmologically as $\lambda\propto a$ and $\nu_\tau\propto a^{-1}$, while in the unit-rate proper-time representation their product remains $\lambda\nu_\tau=c_0$. This illustrates how cosmological evolution of individual dimensional quantities can coexist with exact cancellation in a local physical relation; the corresponding temporal-parametrization dependence is developed explicitly in Sec.~\ref{subsec:photon_temporal_representation}.

By ``cross-epoch observability'' we do not mean that dimensional quantities become separately observable merely because they are evaluated at different epochs. We use the term \emph{cross-epoch observable} for a complete operational relation connecting quantities measured or standardized at different cosmological epochs. A correlated model may assign different slice-wise numerical realizations, but empirical content belongs only to combinations that connect the slices. Each measured dimensional quantity is referred to a locally realized physical reference standard, and the final comparison must be invariant under consistent slice-wise rescalings of both the measured quantities and their corresponding standards.

Let $R_{\alpha,i}$ denote the locally realized reference standard for $Q_{\alpha,i}$, $\mathcal P_{e\rightarrow o}$ the propagation or comparison map, and $\mathcal S$ any additional source-equivalence or source-standardization prescription required to identify intrinsic source quantities between epochs. A general operational template is
\begin{equation}
\mathcal O_{eo}[\mathcal S] = F\!\left( \left\{\frac{Q_{\alpha,e}}{R_{\alpha,e}}\right\}, \left\{\frac{Q_{\beta,o}}{R_{\beta,o}}\right\}, \mathcal P_{e\rightarrow o}; \mathcal S \right) ,
\label{eq:cross_epoch_observable}
\end{equation}
where $e$ and $o$ denote the emission and observation epochs. Thus, the quantity $\mathcal O_{eo}$ is the complete emission-to-observation observable, not an isolated ratio of dimensional quantities at two epochs. It is defined only after the local reference standards, the propagation or comparison map, and, when required, the source-equivalence prescription have been specified.

$R_{\alpha,i}$ may be a clock transition, calibrated spectral scale, ruler, mass standard, or detector standard appropriate to the measurement. The prescription $\mathcal S$ is not itself a locally realized standard; it specifies how intrinsic source quantities or their distributions are identified across hypersurfaces. In CTD, a source timescale tied to local microphysics leads to proper-time source equivalence and the standard cosmological stretching. The alternative GCT source-equivalence prescription permits the endpoint temporal-rate factor to enter the standardized width relation.

Three levels are useful to distinguish. A coordinate or slice-wise model quantity specifies the representation on a given hypersurface. A cross-epoch model factor specifies how such quantities enter the relation between emission and observation. The measured statistic is obtained after the propagation law, reference standards, and any required source-equivalence prescription are specified.

For the cross-epoch comparisons considered here, this distinction leads to three useful categories. In Class~(i), the correlated $L$--$M$--$T$ factor cancels identically in a genuinely local dimensionless combination. For example, the dimensionless gravitational coupling $\alpha_G\equiv\frac{Gm^2}{\hbar c}$ is independent of the common factor $N_i$ under the tabulated scalings, since $G\propto N_i^4$, $m^2\propto N_i^{-4}$, $\hbar\propto N_i^{-1}$, and $c\propto N_i$.

Class~(ii) contains cross-epoch temporal relations that may retain a relative endpoint factor $\mathcal N_{G,e}/\mathcal N_{G,o}$. The corresponding normalized timing factor is derived in Sec.~\ref{subsec:time_intervals}. It is a model factor defined relative to the specified $\mathcal T_G$, not the measured CTD statistic; under a source-equivalence prescription $\mathcal S$, it enters the standardized width relation discussed in Sec.~\ref{subsec:source_equivalence}.

Class~(iii) contains observables that depend on independent dimensionless parameters or non-universal source physics not fixed by the tabulated $L$--$M$--$T$ scaling. A constraint on $\Delta\alpha/\alpha$ is a representative example: transferring it to the parameter $b$ requires the electromagnetic and measurement model rather than dimensional scaling alone. These classes are conceptually distinct, although they need not be observationally separable in a particular probe. CTD below provides an example in which the endpoint temporal-rate contribution is degenerate with population-dependent source evolution.

These categories organize the present analysis by identifying where the common temporal-rate dependence cancels, where it can survive in a cross-epoch relation, and where additional sector-specific physics is required. The source-standardization prescription $\mathcal S$ is an additional condition on the mapping from a model factor to a measured statistic and may enter Classes~(ii) and~(iii).

\section{Robertson--Walker temporal structure and Generalized Cosmological Time}
\label{sec:lapse_to_gct} 

Weyl's postulate supplies the fundamental timelike congruence that orders the homogeneous hypersurfaces of RW spacetime~\cite{Lee:2023bjz,Lee:2024mal}. The geometry does not select a unique rate for the parameter used along that congruence, so the same ordered hypersurfaces admit a class of smooth monotonic temporal parametrizations. This extends the slice-wise viewpoint emphasized in earlier meVSL discussions: quantities may be spatially constant on each homogeneous hypersurface without having identical numerical values on different hypersurfaces~\cite{Lee:2024mal}.

Comoving proper time is the familiar unit-rate member of this class. GCT retains one specified common temporal parameter for cross-epoch comparisons. This section states the relations needed in the main text; the explicit covariant derivation and lapse-variation calculation are given in Appendices~\ref{app:gct_clock_rate} and~\ref{app:lapse_variation}.

\subsection{Robertson--Walker temporal freedom and the unit-rate representation}
\label{subsec:rw_temporal_freedom} 

Before choosing comoving proper time as the temporal coordinate, we write the RW line element as
\begin{equation}
 ds^2=-N^2(t)c_0^2dt^2+a^2(t)\gamma_{ij}dx^i dx^j,  \label{eq:flrw}
\end{equation}
with $N(t)>0$. For a fundamental observer comoving with the RW foliation, Eq.~\eqref{eq:flrw} gives
\begin{equation}
 d\tau=N(t)\,dt.  \label{eq:propertime}
\end{equation}
Choosing comoving proper time as the coordinate gives $t=\tau$ and $N=1$. This is a valid and convenient representation of the same RW spacetime. More generally, under a smooth monotonic reparametrization $t\rightarrow t'=f(t)$,
\begin{equation}
 N'(t')=N(t)\frac{dt}{dt'}.
 \label{eq:main_lapse_transform}
\end{equation}
Because $dt/dt'$ may vary along the congruence, the rescaling factor at one hypersurface need not equal that at another. Thus, the RW lapse freedom is functional rather than a single global normalization freedom. The RW symmetry assumptions admit the proper-time representation, while temporal reparametrization freedom permits the full class of smooth monotonic parametrizations along the same fundamental congruence~\cite{Ellis:1998ct,Ryder2009,Katanaev:2016byi}.

\subsection{Nondynamical lapse and coordinate freedom}
\label{subsec:pure_lapse} 

The lapse in Eq.~\eqref{eq:flrw} is nondynamical. As reviewed in Appendix~\ref{app:lapse_variation}, the reduced gravitational Lagrangian contains no $\dot N$, and variation with respect to $N$ yields the Hamiltonian/Friedmann constraint rather than an independent propagation equation.

Independently, for a vorticity-free fundamental congruence, the Frobenius theorem permits the four-velocity one-form to be written locally as proportional to the gradient of a time function, $U_\mu=-c_0^2 N\,\nabla_\mu t$ in the present SI convention, corresponding to $u_a=-g\nabla_a t$ in geometrized units~\cite{Ellis:1998ct}. RW homogeneity restricts this temporal-rate factor to be spatially constant on each homogeneous slice, so that $N=N(t)$. This condition is spatial, not temporal: in general
\begin{equation}
 \left.N\right|_{\Sigma_1}=N_1=\textrm{const.}_1,  \qquad  \left.N\right|_{\Sigma_2}=N_2=\textrm{const.}_2,  \qquad N_1\neq N_2
 \label{eq:main_slice_lapse_freedom}
\end{equation}
may hold without violating RW homogeneity. A smooth positive $N(t)$ associated only with a temporal coordinate can be absorbed into a redefinition of that coordinate. Unequal coordinate-lapse values on different hypersurfaces are allowed by the geometry but are not, by themselves, observables.

\subsection{GCT temporal parametrization and covariance}
\label{subsec:gct_scalar} 

For any smooth monotonic time function $\mathcal T$ on the same fundamental congruence,
\begin{equation}
 \mathcal N_{\mathcal T} \equiv \left(U^\mu\nabla_\mu\mathcal T\right)^{-1} =\frac{d\tau}{d\mathcal T}, \label{eq:general_relational_rate}
\end{equation}
where $U^\mu$ is tangent to the fundamental timelike congruence and is normalized by $g_{\mu\nu}U^\mu U^\nu=-c_0^2$. Equation~\eqref{eq:general_relational_rate} is a scalar relation for the chosen $\mathcal T$. Covariance fixes its representation under coordinate changes but does not select a preferred temporal parameter~\cite{Ellis:1998ct,Gourgoulhon:2007ue}.

Reparametrization invariance removes physical significance from an arbitrary, unspecified coordinate lapse. Once a temporal parameter $\mathcal T_G$ is specified and coordinates are adapted to it, the same rate is represented by the ordinary RW lapse, $N(\mathcal T_G)=\mathcal N_G$. The relevant distinction is therefore between an arbitrary coordinate choice and a specified temporal parametrization retained across hypersurfaces, not between two different lapse objects.

For the specified $\mathcal T_G$, Eq.~\eqref{eq:general_relational_rate} gives
\begin{equation}
 \mathcal N_G\equiv\left(U^\mu\nabla_\mu\mathcal T_G\right)^{-1} =\frac{d\tau}{d\mathcal T_G},  \qquad  d\tau=\mathcal N_G\,d\mathcal T_G.
 \label{eq:main_NG_definition}
\end{equation}
We refer to $\mathcal N_G$ as the GCT temporal-rate factor when emphasizing its interval-conversion role. For the specified $\mathcal T_G$, it is a temporal-rate scalar; in adapted coordinates,
\begin{equation}
 \mathcal N_G=N(\mathcal T_G).  \label{eq:main_NG_equals_N}
\end{equation}
The unit-rate choice $\mathcal N_G=1$ identifies $\mathcal T_G$ with proper time up to an additive constant. Within the reparametrization-invariant gravity and matter sector considered here, neither this unit-rate representation nor a non-unit realization is dynamically selected by RW geometry or the Hamiltonian constraint. Lapse variation constrains the geometry but supplies no propagation equation selecting $\mathcal N_G(a)$.

Proper time retains its usual local geometrical status as the metric interval along the fundamental congruence. That local status, however, is not a dynamical theorem selecting $(\tau,1)$ as the unique cross-epoch cosmological temporal realization. Conversely, transforming the metric to proper-time coordinates sets the metric lapse to unity in that coordinate representation but does not by itself identify a separately retained scalar time function $\mathcal T_G$ with $\tau$. Neither $\tau$ nor $\mathcal T_G$ is by itself a microscopic clock mechanism. Operational time measurements compare changes in physical systems, while the rates of atomic, nuclear, hydrodynamic, or other local processes are determined by the corresponding interactions and source physics. The temporal variables specify how such intervals are represented; they do not determine the microscopic dynamics.

GCT uses the residual temporal-parametrization freedom left after the fundamental congruence and homogeneous slicing have been fixed. It retains the same specified scalar parameter $\mathcal T_G$ when comparing different hypersurfaces. Once that parameter is specified, changing its rate differently at different epochs requires a nontrivial redefinition $\mathcal T'_G=f(\mathcal T_G)$ and a different temporal parametrization, not merely an overall normalization of the same $\mathcal T_G$. Consequently, the RW geometry does not by itself select between these admissible temporal realizations; their relevance to cross-epoch cosmological comparisons must be determined operationally through complete observables. Appendix~\ref{app:gct_clock_rate} gives the explicit transformation.

\subsection{Photon redshift and the unit-rate representation}
\label{subsec:photon_temporal_representation}

A simple photon example makes the role of the temporal parametrization concrete. For a freely propagating photon in RW spacetime, the physical wavelength measured on the homogeneous hypersurfaces follows the usual cosmological redshift,
\begin{equation}
 \lambda_i=a_i\lambda_0 ,
 \label{eq:main_photon_lambda}
\end{equation}
where the present normalization is $a_0=1$. In the unit-rate proper-time representation, the corresponding frequency scales as
\begin{equation}
 \nu_{\tau,i}=a_i^{-1}\nu_0 ,
 \label{eq:main_photon_nu_tau}
\end{equation}
so that
\begin{equation}
 \lambda_i\nu_{\tau,i}=c_0 .
 \label{eq:main_photon_c_tau}
\end{equation}
Thus, the wavelength and frequency, both dimensionful quantities, evolve between cosmological hypersurfaces even though their product remains the epoch-independent local light speed in the unit-rate representation.

For the same photon phase described using the retained temporal parameter $\mathcal T_G$, the frequency is defined with respect to $\mathcal T_G$. Since $d\tau=\mathcal N_G\,d\mathcal T_G$,
\begin{equation}
 \nu_G\equiv\frac{1}{2\pi}\frac{d\phi}{d\mathcal T_G} = \mathcal N_G\frac{1}{2\pi}\frac{d\phi}{d\tau} = \mathcal N_G\nu_\tau . \label{eq:main_photon_frequency_transform}
\end{equation}
At fixed spatial length standard, the corresponding light-speed numerical realization is 
\begin{equation}
 c_G\equiv\lambda\nu_G=\mathcal N_Gc_0 . \label{eq:main_photon_cG}
\end{equation}
This does not assign independent observability to a coordinate lapse or to the isolated numerical value of a dimensional constant. Rather, it shows that $c_0$ is the unit-rate member of the same temporal-parametrization structure. Choosing $\mathcal N_G=1$ leaves the cosmological evolution of wavelength and frequency intact while their product remains constant.

For a nontrivial GCT realization, homogeneity implies
\begin{equation}
 \left.\mathcal N_G\right|_{\Sigma_i}=N_i=\textrm{const.}_i, \qquad \left.c_G\right|_{\Sigma_i}=c_i=N_i c_0=\textrm{const.}_i , \label{eq:main_slice_c_constant}
\end{equation}
so the light-speed realization is constant over each homogeneous hypersurface. This is a spatial statement, not an equality condition between epochs: for two hypersurfaces $\Sigma_i$ and $\Sigma_j$,
\begin{equation}
 N_i\neq N_j \quad\Longrightarrow\quad c_i\neq c_j  \label{eq:main_cross_slice_c_difference}
\end{equation}
is compatible with the same RW homogeneity and the same Weyl congruence.

For the meVSL realization $\mathcal N_G=a^{b/4}$,
\begin{equation}
 \lambda\propto a,\qquad \nu_G\propto a^{-1+b/4},\qquad c_G=c_0a^{b/4}.  \label{eq:main_photon_mevsl_scaling}
\end{equation}
The distinction is therefore between the evolution of dimensionful wave quantities and the temporal convention used to represent their rates, rather than between a redshifting and a non-redshifting spatial geometry.

\subsection{Emission and observation hypersurfaces}
\label{subsec:cross_epoch} 

For the homogeneous $\mathcal T_G$ parametrization, $\mathcal N_G$ is spatially constant on each constant-$\mathcal T_G$ hypersurface,
\begin{equation}
 \left.\mathcal N_G\right|_{\Sigma_i}=\mathcal N_{G,i}=\textrm{const.}_i,  \label{eq:main_slice_constant}
\end{equation}
but the constants on different hypersurfaces need not coincide. Thus $\mathcal N_{G,e}\neq\mathcal N_{G,o}$ is compatible with the same homogeneous RW geometry. A comparison of the endpoint rates acquires a definite meaning only after a common $\mathcal T_G$ has been specified. We define
\begin{equation}
 \mathcal R_G(e,o)\equiv\frac{\mathcal N_{G,e}}{\mathcal N_{G,o}}.  \label{eq:main_RG}
\end{equation}
The endpoint values refer to different cosmological hypersurfaces and are not simultaneous local measurements. Their relevance arises only through a complete cross-epoch relation. CTD provides the timing comparison considered below.

A useful structural analogy is the familiar distinction between comoving and physical length: on a hypersurface $\Sigma_i$, a fixed comoving interval is converted to a local physical interval by $a_i$. In the temporal sector, a fixed $d\mathcal T_G$ is converted to the corresponding local proper-time interval by $\mathcal N_{G,i}$. The analogy concerns slice-dependent conversion factors only and does not identify the scale factor with the lapse~\cite{Hogg:1999ad,Weinberg2008,Dodelson:2020bqr}.

\section{Cosmological time-dilation relations and source equivalence}
\label{sec:observables} 

We summarize the CTD relation and focus on the source-standardization step relevant to the present observability question. The detailed time-dilation formulation was developed previously in Ref.~\cite{Lee:2026kyz}.

\subsection{Proper-time CTD and the GCT interval relation}
\label{subsec:time_intervals} 

For null-linked emission and observation intervals measured in local proper time, the standard RW result is
\begin{equation}
\Delta\tau_o=(1+z)\Delta\tau_e.  \label{eq:properCTD}
\end{equation}
This relation is independent of the temporal parametrization used to describe the same RW spacetime~\cite{SupernovaCosmologyProject:2001ziv,Lee:2023ucu,DES:2024vgg,Lee:2024kxa}.

Using $d\tau=\mathcal N_G\,d\mathcal T_G$ at the two endpoints gives
\begin{equation}
 \frac{\Delta\mathcal T_{G,o}}{\Delta\mathcal T_{G,e}} = (1+z)\frac{\mathcal N_{G,e}}{\mathcal N_{G,o}}.  \label{eq:TGCTD}
\end{equation}
Here and below, the endpoint relations assume that the source and observer intervals are short compared with the cosmological variation timescale of $\mathcal N_G$, so that the temporal-rate factor may be evaluated on the corresponding endpoint hypersurface. For $\mathcal N_G=a^{b/4}$,
\[
\frac{\Delta\mathcal N_G}{\mathcal N_G}\simeq \frac{b}{4}H\Delta\tau ,
\]
which is negligible for the day-to-year intrinsic timescales relevant to SNe, GRBs, and QSOs variability compared with a Hubble time. Thus, we define the normalized timing ratio
\begin{equation}
 \mathcal R_T(z)\equiv  \frac{\Delta\mathcal T_{G,o}}{(1+z)\Delta\mathcal T_{G,e}}  =\frac{\mathcal N_{G,e}}{\mathcal N_{G,o}}  =\mathcal R_G(e,o). \label{eq:RT}
\end{equation}
Equation~\eqref{eq:RT} compares temporal rates on two distinct homogeneous hypersurfaces with respect to the same specified $\mathcal T_G$. In the meVSL realization, its departure from unity is controlled jointly by $b$ and the separation in scale factor between emission and observation.

A constant rescaling $\mathcal T_G\rightarrow\lambda\mathcal T_G$, with $\lambda>0$, sends $\Delta\mathcal T_G\rightarrow\lambda\Delta\mathcal T_G$ and $\mathcal N_G\rightarrow\mathcal N_G/\lambda$, leaving Eq.~\eqref{eq:RT} unchanged. Such an overall normalization can set one endpoint value, for example $\mathcal N_{G,o}=1$, but cannot remove a nontrivial relative rate $\mathcal N_{G,e}/\mathcal N_{G,o}$ unless the two endpoint values are equal.

By contrast, a general redefinition $\mathcal T'_G=f(\mathcal T_G)$ gives $\mathcal N'_G=\mathcal N_G/f'(\mathcal T_G)$ and may rescale the two endpoints differently. This defines a different temporal parametrization rather than a mere normalization of the retained $\mathcal T_G$. Equation~\eqref{eq:TGCTD} is the proper-time CTD relation expressed in the $\mathcal T_G$ parametrization, not an additional propagation law.

\subsection{Source equivalence in cosmological time-dilation measurements}
\label{subsec:source_equivalence} 

The operational template in Eq.~\eqref{eq:cross_epoch_observable} can be instantiated directly for CTD. For a standardized width or duration, let $R_{T,i}$ denote the locally realized temporal reference standard on $\Sigma_i$, $\mathcal P^{\textrm{null}}_{e\rightarrow o}$ the null-propagation map represented by Eq.~\eqref{eq:properCTD}, and $\mathcal S$ the prescription used to identify the intrinsic source-timescale distribution between epochs. Schematically,
\begin{equation}
\mathcal O^{\textrm{CTD}}_{eo}[\mathcal S] = F_{\textrm{CTD}}\!\left( \frac{W_{\textrm{int},e}}{R_{T,e}}, \frac{W_{\textrm{obs},o}}{R_{T,o}}, \mathcal P^{\textrm{null}}_{e\rightarrow o};
\mathcal S \right). \label{eq:ctd_operational_template}
\end{equation}
Under a consistent slice-wise rescaling, $W_i\rightarrow\xi_i W_i$ and $R_{T,i}\rightarrow\xi_i R_{T,i}$, each ratio in Eq.~\eqref{eq:ctd_operational_template} is unchanged. The nontrivial distinction lies in what the cross-epoch prescription $\mathcal S$ holds fixed.

Astronomical CTD measurements usually compare distinct sources at different redshifts. Such measurements require a source-equivalence condition regardless of the temporal parametrization being tested. We use \emph{source equivalence} for the requirement that, after relevant source properties, rest-frame wavelength dependence, astrophysical evolution, and selection effects have been controlled, the compared sources represent the same intrinsic timescale distribution. \emph{Source standardization} denotes the observational or statistical procedure used to establish or test this condition.

For an intrinsic interval, the two parametrizations are related by
\begin{equation}
 W_{\tau,i}=\mathcal N_{G,i}W_{G,i}.  \label{eq:source_interval_relation}
\end{equation}
Source equivalence may be formulated in GCT intervals,
\begin{equation}
 p_G(W_G\mid\bm\theta,z)=p_G(W_G\mid\bm\theta),  \label{eq:gct_source_equivalence}
\end{equation}
or directly in proper-time intervals,
\begin{equation}
 p_\tau(W_\tau\mid\bm\theta,z)=p_\tau(W_\tau\mid\bm\theta),  \label{eq:proper_source_equivalence}
\end{equation}
where $\bm\theta$ denotes controlled source properties. These are alternative empirical standardization prescriptions. Neither defines $\mathcal T_G$, selects a temporal parametrization, nor modifies Eq.~\eqref{eq:main_NG_definition}. The temporal realization $(\mathcal T_G,\mathcal N_G)$ and the source-equivalence prescription are logically separate.

The role of $\mathcal S$ is operational: it specifies how standardized source populations are identified between epochs after the relevant source physics and observational corrections have been modeled. It does not assert that either $\tau$ or $\mathcal T_G$ controls the microscopic rate of the source process.

When the standardized source duration is tied to the same local microphysics, the tabulated $L$--$M$--$T$ scaling law favors the proper-time form. A local period $P$ has time dimension, so its numerical value in the slice-wise time unit scales as $P_G\propto N_i^{-1}$. A source duration that is a fixed multiple of such a period then satisfies
\[
W_G\propto N_i^{-1}, \qquad W_\tau=N_i W_G=\mathrm{const.},
\]
which is Eq.~\eqref{eq:proper_source_equivalence}. By contrast, Eq.~\eqref{eq:gct_source_equivalence} requires the intrinsic $\mathcal T_G$ interval itself to be epoch independent. This is an alternative cross-epoch source-equivalence prescription, not a statement about the microscopic mechanism producing the source timescale. Its adequacy must be tested together with source-evolution and selection models. The prescription is applied separately to each source class and does not assume a common microscopic origin for the intrinsic timescales of SNe, GRBs, QSOs, or other populations.

In proper-time language, the GCT source-equivalence prescription corresponds to
\[
W_{\tau,\mathrm{int}}(a)\propto \mathcal N_G(a),
\]
whereas a microphysics-tied intrinsic source timescale in the tabulated sector gives $W_{\tau,\mathrm{int}}(a)=\mathrm{const.}$ For $\mathcal N_G(a)=a^{b/4}$, the former becomes $W_{\tau,\mathrm{int}}\propto a^{b/4}$. To parameterize residual astrophysical evolution relative to that GCT source-equivalence prescription, we write
\begin{equation}
 W_G\propto a^{\eta_{\mathrm{astro}}},  \label{eq:source_evolution_eta}
\end{equation}
and denote by $W_{\mathrm{obs}}$ the standardized observer-frame width or duration. The observed stretching then becomes
\begin{equation}
W_{\mathrm{obs}}\propto (1+z)^{1-b/4-\eta_{\mathrm{astro}}}, \qquad n=1-\frac{b}{4}-\eta_{\mathrm{astro}} , \label{eq:ctd_eta_degeneracy}
\end{equation}
where $\eta_{\mathrm{astro}}$ is a phenomenological source-evolution exponent defined relative to the adopted GCT source-equivalence prescription. It summarizes, at the level of one bookkeeping parameter, the source-dependent intrinsic-duration effects discussed previously for SNe, GRBs, and QSOs~\cite{Lee:2026kyz}.

If the same residual evolution is instead parameterized directly in proper time by $W_{\tau,\mathrm{int}}\propto a^{\eta_\tau}$, then $W_\tau=\mathcal N_GW_G$ gives
\begin{equation}
\eta_\tau=\eta_{\mathrm{astro}}+\frac{b}{4}. \label{eq:eta_proper_conversion}
\end{equation}
A source-evolution constraint quoted in proper-time units cannot be inserted directly as $\eta_{\mathrm{astro}}$ without this conversion. For a specified $\mathcal T_G$ and the GCT source-equivalence prescription, the $b/4$ term is associated with the endpoint temporal-rate ratio, while $\eta_{\mathrm{astro}}$ parameterizes residual source evolution. The fitted CTD exponent constrains the combination $\eta_\tau=b/4+\eta_{\mathrm{astro}}$ unless additional information separates the two contributions.

For several independently standardized source populations $X$, we write
\begin{equation}
n_X=1-\frac{b}{4}-\eta_{\mathrm{astro},X},
\label{eq:ctd_multipop}
\end{equation}
where $b$ is common to the adopted GCT/meVSL realization, while source physics and observational systematics are population dependent. If every residual source term is represented by an unconstrained power law with the same redshift dependence, a common-shift degeneracy remains.

In an actual analysis, source evolution, wavelength dependence, calibration, selection effects, and other nuisance contributions are modeled for each population and need not share the same functional form. Separation may come from population-specific source models, independent constraints or priors on source evolution, nuisance terms with a different redshift dependence, or a joint cross-population likelihood in which $b$ is common while the astrophysical parameters are not. Extracting a common $b$ is an observational-inference problem: independent analyses of SNe, GRBs, QSOs, and other source classes test whether a common contribution remains after their distinct astrophysical and instrumental effects are controlled. Conventional SNe~Ia analyses likewise require extensive standardization of light-curve properties, host correlations, source evolution, and selection effects~\cite{SNLS:2007cqk,SNLS:2010kps,Childress:2014vka,Kessler:2016uwi,Brout:2020msh,Kenworthy:2021azy,Brout:2022vxf}.

For the observational stretching law $(1+z)^n$, the Dark Energy Survey analysis found $n=1.003\pm0.005\,(\textrm{stat})\pm0.010\,(\textrm{sys})$ from the combined bands, consistent with $n=1$~\cite{DES:2024vgg}. A band-wise analysis of the DES data reported $n=0.988\pm0.008\,(\textrm{stat})\pm0.010\,(\textrm{sys})$ in the $i$ band~\cite{Lee:2024kxa}. Earlier meVSL studies used measured CTD exponents to constrain the corresponding model parameter~\cite{Lee:2023ucu,Lee:2024kxa}. In the present formulation, those results correspond to the special case $\eta_{\mathrm{astro}}=0$ under the GCT source-equivalence prescription. More generally, the same CTD measurements constrain $b/4+\eta_{\mathrm{astro},X}$ for each source class, while under proper-time source equivalence they do not by themselves constrain $b$.

If source equivalence is formulated for $W_G$, Eq.~\eqref{eq:TGCTD} gives
\begin{equation}
 W_{G,o} =  (1+z)\frac{\mathcal N_{G,e}}{\mathcal N_{G,o}}W_{G,e}.  \label{eq:gct_population}
\end{equation}
For $\mathcal N_G(a)=a^{b/4}$, with $a_o=1$ and $a_e=(1+z)^{-1}$,
\begin{equation}
 \frac{W_{G,o}}{W_{G,e}}=(1+z)^{1-b/4},  \qquad  b=4(1-n), \qquad (\eta_{\mathrm{astro}}=0).  \label{eq:population_powerlaw}
\end{equation}
If source equivalence is instead formulated in proper time, Eq.~\eqref{eq:properCTD} is recovered. A CTD inference of $b$ is conditional on the adopted source-equivalence, source model, and observational treatment. The role of the observations is to separate any common temporal-rate contribution from population-specific astrophysical and instrumental effects within the full likelihood. An inferred $b\neq0$ under the GCT prescription would indicate a non-unit endpoint-rate scaling within this parametrization, but would not, by itself, establish the evolution of any individual dimensional quantity.

\section{Phenomenological parametrization and scope}
\label{sec:phenomenology} 

We now state the phenomenological temporal rate used in meVSL and separate the CTD test from the broader correlated dimensional sector. The ``minimally extended'' designation refers to retaining the standard local form of the physical relations considered in the framework on each homogeneous hypersurface, while allowing their slice-wise numerical realizations to differ across cosmological epochs in a correlated manner. The correlated dimensional sector has been developed in earlier meVSL studies and is summarized here only to clarify its relation to dimensional-constant arguments.

\subsection{Phenomenological meVSL realization}
\label{subsec:minimal_realization} 

At the phenomenological level,
\begin{equation}
 \mathcal N_G(a)=f(a), \qquad f(a)>0, \qquad f(1)=1, \label{eq:general_f}
\end{equation}
with $a_o=1$ and $\mathcal N_{G,o}=1$ fixing the present normalization. meVSL adopts
\begin{equation}
\mathcal N_G(a)=a^{b/4}.  \label{eq:ansatz}
\end{equation}
Along the fundamental congruence, this is equivalent to
\begin{equation}
 d\mathcal T_G=a^{-b/4}d\tau , \label{eq:TG_mevsl_rate}
\end{equation}
which specifies the retained temporal parametrization as a function of the cosmological scale factor without introducing a new propagating clock field. The unit-rate case is $b=0$. At the emission and observation endpoints,
\begin{equation}
 \mathcal R_T(z)=\mathcal R_G(e,o)=(1+z)^{-b/4}. \label{eq:RTb}
\end{equation}
This is the parametrization used in the earlier meVSL program~\cite{Lee:2024zcu,Lee:2023bjz,Lee:2024mal,Lee:2025osx,Lee:2020zts}; we reformulate its temporal meaning rather than introducing a new $b$-dependent law.

\subsection{Observable scope and transfer of constraints}
\label{subsec:scope} 

If $N$ denotes only the lapse associated with an arbitrary temporal coordinate, neither a non-unit lapse coefficient nor an isolated relation such as $\widetilde c=Nc_0$ is by itself a coordinate-independent observable. In the present framework, $\mathcal N_G$ is defined relative to the specified common $\mathcal T_G$. This definition does not make $\mathcal N_G$ separately observable; any empirical content must arise through complete cross-epoch relations involving the same temporal parametrization. The meVSL framework specifies correlated slice-wise numerical realizations of dimensional quantities, summarized in Sec.~\ref{subsec:dimensional_argument}. Those relations are phenomenological model assignments and are not consequences of covariance alone.

The observable content must be derived from the complete combination of quantities entering a specified measurement. Bounds conventionally reported as $\dot G/G$ cannot in general be mapped directly onto the meVSL parameter $b$ without rederiving the measured observable under the same gravitational, matter-sector, and temporal assumptions. A bound on $\dot\alpha/\alpha$ constrains a dimensionless combination, but its mapping to the underlying dimensional scalings likewise depends on the physical system and measurement procedure~\cite{Uzan:2002vq}.

For two hypersurfaces in the meVSL realization,
\begin{equation}
 \frac{c_i}{c_j} = \frac{N_i}{N_j} = \left(\frac{a_i}{a_j}\right)^{b/4},  \label{eq:main_c_endpoint_ratio}
\end{equation}
and thus
\begin{equation}
 \ln\!\left(\frac{c_i}{c_j}\right) = \frac{b}{4}\ln\!\left(\frac{a_i}{a_j}\right).  \label{eq:main_c_log_leverage}
\end{equation}
These equations characterize the model's endpoint scaling; the dimensional ratio is not, by itself, an observable. An empirical test requires a complete cross-epoch relation in which the same endpoint factor survives after local standards and propagation effects have been treated consistently. For small $|b|$, sensitivity to a non-unit realization is correspondingly weak unless either $|b|$ lies within the observational reach or the compared epochs provide a sufficiently large scale-factor lever arm, $|\ln(a_i/a_j)|$.

CTD is the explicit cross-epoch timing application and provides one possible realization of the required lever arm. It constrains the temporal relation in Eqs.~\eqref{eq:TGCTD}--\eqref{eq:population_powerlaw}, not the separate evolution of $c$, $G$, $\hbar$, or another dimensional quantity. With $a_o=1$ and $a_e=(1+z)^{-1}$, the endpoint temporal-rate factor is
\begin{equation}
 \frac{\mathcal N_{G,e}}{\mathcal N_{G,o}} =(1+z)^{-b/4},  \label{eq:main_ctd_endpoint_leverage}
\end{equation}
so higher-redshift sources increase the separation between the compared hypersurfaces. The correlated dimensional sector enters the CTD discussion only through the shared phenomenological parameter $b$.

The propagation and endpoint relations in Eqs.~\eqref{eq:TGCTD}--\eqref{eq:population_powerlaw} do not require the separate $c$, $G$, $\hbar$, or $m$ rows of Table~\ref{tab:correlated_scalings}. The argument that a source timescale tied to local microphysics selects proper-time source equivalence does, however, use the general time-unit scaling $[T]_i=N_i[T]_0$ summarized in Sec.~\ref{subsec:dimensional_argument}. Practical inference remains conditional on source standardization and source evolution, as described in Sec.~\ref{subsec:source_equivalence}.

Tests of the broader meVSL sector require dedicated model-specific derivations. No microscopic or dynamical principle is assumed here to select a particular $\mathcal N_G(a)$. The empirical question is whether specified observables, under their required standardization and source-model assumptions, are consistent with the unit-rate case or favor a non-unit temporal-rate realization within the stated parametrization.

\subsection{Correlated meVSL scalings and the Duff distinction}
\label{subsec:dimensional_argument}

The correlated meVSL sector is interpreted under the unit-dependence premise of Sec.~\ref{sec:duff_cross_epoch}: it does not assign independent observability to $c$, $G$, $\hbar$, or $m$~\cite{Duff:2001ba,Duff:2002vp}. Its role is to determine which complete combinations cancel or retain the common temporal-rate dependence in a specified cross-epoch measurement.

Within meVSL, the tabulated dimensional quantities are assigned correlated slice-wise numerical realizations so that the corresponding local defining relations retain their standard form on each homogeneous hypersurface. The detailed derivations belong to the earlier meVSL literature~\cite{Lee:2024zcu,Lee:2023bjz,Lee:2024mal,Lee:2025osx,Lee:2020zts}; Table~\ref{tab:correlated_scalings} collects the relations needed for the present discussion.

The local consistency is tied directly to the slice-wise constancy of the GCT temporal rate. On a fixed homogeneous hypersurface $\Sigma_i$,
\begin{equation}
 \mathcal N_G|_{\Sigma_i}=N_i=\textrm{const.}_i, \qquad c_i=N_ic_0=\textrm{const.}_i .  \label{eq:main_local_slice_constants}
\end{equation}
Local experiments on that hypersurface involve fixed numerical values of the relevant dimensional quantities. The correlated meVSL assignments preserve the standard local form of relations such as
\begin{equation}
 m_i c_i^2=m_0c_0^2,\qquad  \hbar_i c_i=\hbar_0c_0,\qquad  \frac{G_i}{c_i^4}=\frac{G_0}{c_0^4}.  \label{eq:main_local_law_preservation}
\end{equation}
The model does not require a local observer on $\Sigma_i$ to measure a continuously varying $c$ within that slice. The nontrivial structure enters only when quantities standardized on distinct hypersurfaces, for which $N_i$ and $N_j$ may differ, are compared within one specified cross-epoch construction.

The correlated scalings in this sector have a simple algebraic consequence. At the level of the numerical realization on a single homogeneous slice, they are equivalent to the base-unit rescaling
\[
[L]_i=[L]_0,\qquad [T]_i=N_i[T]_0,\qquad [M]_i=N_i^2[M]_0.
\]
If a quantity $Q$ has dimensions
\[
[Q]=L^pM^qT^r,
\]
then the slice-wise numerical realization generated solely by this scaling is
\[
Q_i=N_i^{-(2q+r)}Q_0.
\]
This reproduces $c_i=N_ic_0$, $\hbar_i=N_i^{-1}\hbar_0$, $m_i=N_i^{-2}m_0$, and $G_i=N_i^4G_0$. The statement applies to the tabulated $L$--$M$--$T$ sector with independent dimensionless couplings and mass ratios held fixed.

It follows immediately that every purely local dimensionless monomial constructed entirely within this sector is independent of $N_i$: once the net powers of $L$, $M$, and $T$ vanish, so does the net power of $N_i$. More generally, any quantity satisfying $2q+r=0$ is $N_i$-independent even if it remains dimensionful. The same bookkeeping is consistent with the GCT interval relation. Let $[T]_0$ denote a fixed reference time unit and define $[T]_i=N_i[T]_0$ as the slice-wise bookkeeping time unit associated with the correlated dimensional scaling. This unit language is an algebraic representation of the same temporal-rate factor $N_i$ and does not introduce an additional locally realized clock standard. The physical temporal reference standard entering the CTD observable is $R_{T,i}$ in Eq.~\eqref{eq:ctd_operational_template}.

Representing the same interval in the two conventions gives $d\tau=N_i\,d\mathcal T_G$. The covariant definition $\mathcal N_G=(U^\mu\nabla_\mu\mathcal T_G)^{-1}$ remains the geometric definition of the retained scalar temporal parameter, while the unit-rescaling picture supplies the corresponding algebraic realization of the tabulated dimensional sector.

Let $N_i\equiv\mathcal N_{G,i}$ on a homogeneous hypersurface $\Sigma_i$. In meVSL, $N_i=a_i^{b/4}$. The listed quantities are slice-wise model realizations associated with the retained GCT parametrization. They are not, individually, claims of directly observable local proper-time quantities; empirical meaning belongs to the complete relation in which they enter.

\begin{table*}[t]
\centering
\small
\caption{Representative correlated relations in the meVSL realization, summarized from earlier derivations. The slice-wise entries are model assignments in the retained GCT parametrization and are not separately observable. The third column displays combinations relevant to local or cross-epoch relations, in which common temporal-rate factors may cancel or survive.}
\label{tab:correlated_scalings}
\begin{tblr}{
  width=\textwidth,
  colspec={X[0.18,l] X[0.26,l] X[0.56,l]},
  row{1}={font=\bfseries},
  rows={valign=t},
  stretch=1.08,
}
\toprule
Sector & Slice-wise realization & Local or cross-epoch relation \\
\midrule
Temporal rate & $N_i=a_i^{b/4}$ & $d\tau=N_i\,d\mathcal T_G$; $N_i$ is spatially constant on $\Sigma_i$. \\
Light-speed realization & $c_i=N_i c_0$ & Used with the wave relation $c_i=\lambda_i\nu_{G,i}$ in the retained temporal convention. \\
Gravitation & $G_i=N_i^4G_0$ & $8\pi G_i/c_i^4=8\pi G_0/c_0^4$; the Einstein coupling coefficient is unchanged. \\
Massive local state & $m_i=N_i^{-2}m_0$ & $m_i c_i^2=m_0c_0^2$ for an equivalent local massive state. \\
Quantum sector & $\hbar_i=N_i^{-1}\hbar_0$ & $\hbar_i c_i=\hbar_0c_0$. \\
Propagating photon & $\lambda_i=a_i\lambda_0$, $\nu_{G,i}=N_i a_i^{-1}\nu_0$ & $c_i=\lambda_i\nu_{G,i}=N_ic_0$ in the retained temporal convention; $h_i\equiv2\pi\hbar_i$ and $h_i\nu_{G,i}=h_0\nu_0 a_i^{-1}$; see Sec.~\ref{subsec:photon_temporal_representation}. \\
\bottomrule
\end{tblr}
\end{table*}

The table realizes the Class~(i) cancellation described in Sec.~\ref{sec:duff_cross_epoch}. In particular, $G_i/c_i^4$, $\hbar_i c_i$, and $m_i c_i^2$ retain their corresponding local forms, while the photon-energy relation keeps the physical factor $a_i^{-1}$ after the common $N_i$ factors cancel. The CTD construction of Sec.~\ref{sec:observables} provides the Class~(ii) example. Proper-time source equivalence gives the standard $n=1$ stretching, whereas the GCT source-equivalence prescription can retain the endpoint temporal-rate factor; with residual source evolution, the fitted exponent depends on $b/4+\eta_{\mathrm{astro}}$. Class~(iii) lies outside the universal $L$--$M$--$T$ rescaling and requires independent dimensionless or model-specific input, as in transferring a reported constraint on $\Delta\alpha/\alpha$ to $b$.

\section{Conclusion}
\label{sec:conclusion}

The unit-dependence argument remains intact: an isolated numerical change of a dimensional quantity is not by itself a unit-independent observable. The distinct question addressed here is whether a complete relation between standards realized at different cosmological epochs retains observable content after the local standards, propagation or comparison map, and any required cross-epoch identification prescription are specified together.

The Robertson--Walker temporal sector provides the relevant setting. The cosmological principle and Weyl's postulate determine the homogeneous and isotropic geometry and its fundamental congruence without uniquely fixing the temporal rate along it. Generalized cosmological time retains one specified parameter $\mathcal T_G$ from this class, with $d\tau=\mathcal N_Gd\mathcal T_G$, and $(\tau,1)$ is the unit-rate member. Within the reparametrization-invariant sector, neither this unit-rate representation nor a non-unit realization is dynamically selected by RW geometry or the Hamiltonian constraint. An arbitrary coordinate lapse remains gauge, while physical clocks and source timescales are supplied by the relevant local dynamics rather than by the temporal parametrization itself.

In the minimally extended varying speed of light realization, $\mathcal N_G(a)=a^{b/4}$ and the tabulated $L$--$M$--$T$ quantities are correlated through the same parameter $b$. Their slice-wise assignments preserve the standard local form of the physical relations considered in the framework. Purely local dimensionless combinations cancel the common $N_i$ dependence identically. Temporal-rate dependence can therefore enter an empirical prediction only through additional cross-epoch comparison structure or through dimensionless/model-specific physics not fixed by the correlated $L$--$M$--$T$ scaling.

Cosmological time dilation realizes the first possibility. The proper-time propagation law remains $\Delta\tau_o=(1+z)\Delta\tau_e$. A source duration tied to the same local microphysics selects proper-time source equivalence and gives the standard $n=1$ stretching. Under the alternative GCT source-equivalence prescription, $W_G$ is standardized across epochs and the meVSL power law gives $n=1-b/4-\eta_{\textrm{astro}}$. Equivalently, $\eta_\tau=b/4+\eta_{\textrm{astro}}$, so the fitted exponent constrains a composite temporal-rate and source-evolution contribution unless independent information separates them. The cosmological parameter $b$ is common to the adopted realization, whereas source evolution and instrumental systematics are population dependent.

For the meVSL power law, the endpoint departure scales as $(b/4)\ln(a_i/a_j)$, so detectability requires both adequate parameter sensitivity and sufficient cosmological leverage. Compatibility with $b=0$ or $\mathcal N_G=1$ does not by itself establish the proper-time representation as a uniquely preferred cross-epoch realization. Empirical discrimination applies to complete observable relations constructed from a specified temporal parametrization together with the required standardization assumptions, rather than to a coordinate gauge by itself. Conversely, under proper-time source equivalence, increased CTD precision alone does not generate sensitivity to $b$. A stronger empirical case for a non-unit realization would require a consistent temporal-rate contribution across independent observables with distinct astrophysical and instrumental systematics. Thus, the unit dependence of dimensional quantities does not by itself settle the observational content of complete cross-epoch cosmological relations.

\section*{Acknowledgements}
SL is supported by the Basic Science Research Program through the National Research Foundation of Korea (NRF), funded by the Ministry of Science and ICT under Grant No. NRF-2022R1A2C1005050.

\appendix

\section{Temporal-rate relation, temporal reparametrization, and the Robertson--Walker lapse}
\label{app:gct_clock_rate}

This Appendix gives the covariant derivation of the temporal-rate relation used in the main text. We distinguish the RW temporal reparametrization freedom, the representation of that relation in coordinates adapted to $\mathcal T_G$, and the phenomenological temporal-rate function adopted in the meVSL realization. The construction introduces no additional propagating scalar degree of freedom. The common cosmological time parameter $\mathcal T_G$ belongs to the class of temporal parametrizations admitted along the RW fundamental congruence.

\subsection{General RW lapse and proper time along the fundamental congruence}
\label{app:metric_lapse_vs_NG} 

We use the temporal-coordinate convention
\begin{equation}
x^0\equiv c_0t. \label{eq:app_x0_convention}
\end{equation}
Before choosing proper time along the fundamental congruence as the temporal coordinate, the RW metric may be written as
\begin{equation}
ds^2 = -N^2(t)(dx^0)^2 + a^2(t)\gamma_{ij}dx^i dx^j = -N^2(t)c_0^2dt^2 + a^2(t)\gamma_{ij}dx^i dx^j , \label{eq:app_flrw_lapse}
\end{equation}
where $N(t)>0$ and $t$ is a smooth monotonic parameter along the fundamental congruence~\cite{Ellis:1998ct,Gourgoulhon:2007ue}.

For an observer following the fundamental congruence, $dx^i=0$, and the corresponding proper-time interval is
\begin{equation}
d\tau=N(t)\,dt, \qquad N(t)=\frac{d\tau}{dt}. \label{eq:app_proper_metric_lapse}
\end{equation}
With $x^0=c_0t$, the four-velocity field tangent to the fundamental timelike congruence is
\begin{equation}
U^\mu = \left( \frac{c_0}{N(t)},0,0,0 \right), \label{eq:app_comoving_four_velocity}
\end{equation}
which satisfies
\begin{equation}
g_{\mu\nu}U^\mu U^\nu=-c_0^2.
\end{equation}

Under a smooth, monotonically increasing temporal reparametrization $t\rightarrow t'=f(t)$, the lapse transforms as
\begin{equation}
N'(t') = N(t)\frac{dt}{dt'}. \label{eq:app_lapse_transform}
\end{equation}
Because $dt/dt'$ is generally a function of time, the rescaling applied at two hypersurfaces need not be the same. The lapse freedom is functional, not merely an overall constant rescaling. The explicit lapse function depends on the chosen temporal coordinate. In particular, choosing proper time along the fundamental congruence as the temporal coordinate gives the unit-lapse representation without changing the underlying RW spacetime.

\subsection{Covariance and temporal-parametrization freedom} 

Let $\mathcal T$ be any smooth, monotonically increasing temporal parameter that is constant on the homogeneous spatial slices and parametrizes the same fundamental congruence. The temporal-rate relation may be written covariantly as
\begin{equation}
\mathcal N_{\mathcal T} \equiv \left(U^\mu\nabla_\mu\mathcal T\right)^{-1} = \frac{d\tau}{d\mathcal T}. \label{eq:app_general_scalar_rate}
\end{equation}
For the coordinate time $t$, this relation becomes
\begin{equation}
\left(U^\mu\nabla_\mu t\right)^{-1} = \frac{d\tau}{dt} = N(t). \label{eq:app_t_reproduces_lapse}
\end{equation}

Thus, the covariant temporal-rate relation is available for the class of smooth monotonic temporal parametrizations along the fundamental congruence. Covariance ensures the consistent representation of this relation under coordinate transformations, but it does not select a preferred temporal parameter.

Within this class, GCT keeps explicit a common $\mathcal T_G$ and defines
\begin{equation}
\mathcal N_G
\equiv
\left(U^\mu\nabla_\mu\mathcal T_G\right)^{-1}
=
\frac{d\tau}{d\mathcal T_G}.
\label{eq:app_NG_definition}
\end{equation}
Accordingly,
\begin{equation}
d\tau
=
\mathcal N_G\,d\mathcal T_G.
\label{eq:app_NG_rate}
\end{equation}

Equation~\eqref{eq:app_NG_rate} is the SI-unit form of the standard hypersurface-orthogonal time-function relation. For a vorticity-free fundamental congruence, the Frobenius theorem allows the four-velocity one-form to be written locally as
\begin{equation}
U_\mu = -c_0^2\,\mathcal N_{\mathcal T}\nabla_\mu\mathcal T, \label{eq:app_frobenius_temporal_rate}
\end{equation}
the SI counterpart of the familiar relation $u_a=-g\nabla_a t$ used in covariant cosmology~\cite{Ellis:1998ct}. Contracting Eq.~\eqref{eq:app_frobenius_temporal_rate} with $U^\mu$ and using $g_{\mu\nu}U^\mu U^\nu=-c_0^2$ reproduces Eq.~\eqref{eq:app_general_scalar_rate}. Equation~\eqref{eq:app_NG_rate} is the standard hypersurface-orthogonal temporal-rate relation evaluated for the specified $\mathcal T_G$.

It is not an additional propagation law or a new dynamical equation. The proper-time representation is the unit-rate member of this class: $\mathcal N_G=1$ gives $\mathcal T_G=\tau+\textrm{const.}$, or $\mathcal T_G=\tau$ when their temporal origins are synchronized.

\subsection{Coordinates adapted to \texorpdfstring{$\mathcal T_G$}{TG}} 

Define
\begin{equation}
x_G^0\equiv c_0\mathcal T_G.
\end{equation}
In coordinates adapted to the common $\mathcal T_G$, the RW metric takes the form
\begin{equation}
ds^2 = -N^2(\mathcal T_G)c_0^2d\mathcal T_G^2 + a^2(\mathcal T_G)\gamma_{ij}dx^i dx^j. \label{eq:app_metric_TG}
\end{equation}
For an observer following the fundamental congruence,
\begin{equation}
d\tau = N(\mathcal T_G)\,d\mathcal T_G.
\end{equation}
Comparison with Eq.~\eqref{eq:app_NG_rate} gives
\begin{equation}
\mathcal N_G=N(\mathcal T_G). \label{eq:app_NG_equals_N}
\end{equation}

For a specified common $\mathcal T_G$, $\mathcal N_G$ is the corresponding temporal-rate scalar; in coordinates adapted to $\mathcal T_G$, it is represented by the ordinary RW lapse.  GCT retains the same $\mathcal T_G$ explicitly throughout the cross-epoch temporal relations rather than fixing the lapse to unity from the outset.

A transformation to proper-time coordinates gives
\begin{equation}
ds^2 = -c_0^2d\tau^2 + a^2(\tau)\gamma_{ij}dx^i dx^j.
\end{equation}
The metric lapse in these coordinates is unity. However, if the same scalar time function $\mathcal T_G$ is retained, its relation to proper time remains
\begin{equation}
\frac{d\mathcal T_G}{d\tau} = \mathcal N_G^{-1}. \label{eq:app_TG_in_proper_coordinates}
\end{equation}

Thus, choosing proper-time coordinates does not by itself impose $\mathcal N_G=1$ for a retained $\mathcal T_G$. The latter condition additionally identifies $\mathcal T_G$ with $\tau$, up to an additive constant. These statements distinguish a coordinate representation with unit metric lapse from the unit-rate member of the temporal-parametrization class.

Under a constant rescaling $\mathcal T_G\rightarrow\lambda\mathcal T_G$, with $\lambda>0$, the temporal-rate scalar transforms as $\mathcal N_G\rightarrow\mathcal N_G/\lambda$. Hence, for any two endpoints $e$ and $o$,
\begin{equation}
\frac{\mathcal N'_{G,e}}{\mathcal N'_{G,o}} = \frac{\mathcal N_{G,e}}{\mathcal N_{G,o}}. \label{eq:app_constant_rescaling_ratio}
\end{equation}
An overall normalization of the specified $\mathcal T_G$ may set one endpoint value, such as $\mathcal N_{G,o}=1$, but it cannot remove a nontrivial relative endpoint rate.

Under a general reparametrization $\mathcal T'_G=f(\mathcal T_G)$,
\begin{equation}
 \mathcal N'_{G,i}=\frac{\mathcal N_{G,i}}{f'(\mathcal T_{G,i})},  \qquad i\in\{e,o\},  \label{eq:app_general_reparam_endpoint}
\end{equation}
so different endpoint factors are allowed when $f'$ differs between the two hypersurfaces. Such a transformation defines a different temporal parameter rather than a different normalization of the same specified $\mathcal T_G$.

\subsection{Phenomenological meVSL temporal parametrization} 

The general temporal-rate relation does not determine a particular function $\mathcal N_G(a)$. The meVSL realization adopts the phenomenological parametrization
\begin{equation}
\mathcal N_G(a)=a^{b/4}, \label{eq:app_GCT_clock_law}
\end{equation}
which can equivalently be written as
\begin{equation}
\left(U^\mu\nabla_\mu\mathcal T_G\right)^{-1} = a^{b/4}. \label{eq:app_GCT_covariant_law}
\end{equation}
In coordinates adapted to $\mathcal T_G$, this relation is represented by $N(\mathcal T_G)=a^{b/4}$. It is a phenomenological temporal parametrization, not a solution of an independent propagation equation for the lapse.

The same parameter $b$ also appears in the correlated meVSL scalings summarized in Table~\ref{tab:correlated_scalings}. Those relations are phenomenological model assignments, not consequences of covariance, and are not rederived in this Appendix.

\subsection{Emission and observation values} 

The homogeneous RW parametrization requires $\mathcal T_G$ and $\mathcal N_G$ to be spatially homogeneous. Consequently, the temporal rate is spatially constant on each constant-$\mathcal T_G$ hypersurface,
\begin{equation}
\left.\mathcal N_G\right|_{\Sigma_i} = \mathcal N_{G,i} = \textrm{const.}_i.
\end{equation}
In particular,
\begin{equation}
\mathcal N_{G,e} = \textrm{const.}_e, \qquad \mathcal N_{G,o} = \textrm{const.}_o,
\end{equation}
but spatial constancy on each hypersurface does not require equality between different epochs.

For the meVSL power-law parametrization,
\begin{equation}
\mathcal N_{G,e}=a_e^{b/4}, \qquad \mathcal N_{G,o}=a_o^{b/4}.
\end{equation}
Using the present normalization $a_o=1$ and $a_e=(1+z)^{-1}$ gives
\begin{equation}
\mathcal N_{G,o}=1, \qquad \mathcal N_{G,e}=(1+z)^{-b/4}.
\end{equation}
Thus, the observer endpoint is normalized to unity, while the emission endpoint generally differs from unity when $b\neq0$ and $z\neq0$.

Their ratio is
\begin{equation}
\mathcal R_G(e,o) = \frac{\mathcal N_{G,e}}{\mathcal N_{G,o}} = (1+z)^{-b/4}.
\end{equation}
This is the endpoint-rate relation in the specified GCT parametrization. As shown in Sec.~\ref{sec:observables}, the same relation appears when null-linked temporal intervals are expressed in $\mathcal T_G$. Inferring it from comparisons of distinct astrophysical sources additionally requires the source-equivalence condition developed in that section. Such a condition is necessary for the astrophysical comparison, not for the definition of $\mathcal T_G$ or $\mathcal N_G$.

\section{Lapse variation and the Hamiltonian constraint} 
\label{app:lapse_variation} 

For completeness, we summarize the lapse-variation structure of the standard Einstein--Hilbert gravitational action. This calculation establishes the nondynamical character of the RW lapse. The correlated dimensional scalings are specified separately in the meVSL framework~\cite{Gourgoulhon:2007ue,Arnowitt:1962hi}.

Consider
\begin{equation}
S_g = \frac{1}{2\kappa_0} \int dt\,d^3x\, \sqrt{-g}\, (R-2\Lambda), \label{eq:app_EH_action}
\end{equation}
where $\kappa_0$ denotes the conventional gravitational coupling for the metric and coordinate conventions used here. In coordinates $(t,x^i)$, the determinant of Eq.~\eqref{eq:app_flrw_lapse} is
\begin{equation}
\sqrt{-g} = Nc_0a^3\sqrt{\gamma}. \label{eq:app_metric_determinant}
\end{equation}

After removing the standard second-time-derivative boundary term, the reduced gravitational Lagrangian is
\begin{equation}
L_g = \mathcal C_\gamma \left[ -\frac{a\dot a^2}{Nc_0} + kaNc_0 - \frac{\Lambda}{3}a^3Nc_0 \right], \label{eq:app_reduced_lagrangian}
\end{equation}
where $\mathcal C_\gamma$ is a constant containing the gravitational coupling and the comoving spatial-volume factor. The spatial curvature is normalized by the convention used for $\gamma_{ij}$.

The reduced Lagrangian contains no time derivative of the lapse. Its conjugate momentum vanishes,
\begin{equation}
p_N \equiv \frac{\partial L_g}{\partial\dot N} = 0. \label{eq:app_no_lapse_kinetic}
\end{equation}
Variation of the complete action, including matter, with respect to $N$ gives
\begin{equation}
\frac{\delta(S_g+S_m)}{\delta N} = 0,
\end{equation}
which yields the Hamiltonian/Friedmann constraint rather than an independent propagation equation for $N$.

This is the sense in which the RW lapse is nondynamical. For the reduced variational derivation used here, a temporal parametrization such as
\begin{equation}
N(\mathcal T_G)=a^{b/4} \label{eq:app_lapse_after_variation}
\end{equation}
is imposed after variation with respect to the lapse, so that the Hamiltonian constraint is retained explicitly. Substituting this relation into the action before varying with respect to the lapse would restrict the allowed variations prematurely and obscure the constraint structure.

The non-unit-lapse parametrization does not introduce a new propagating lapse degree of freedom. Within GCT, $\mathcal T_G$ is retained explicitly, and the meVSL realization adopts its temporal-rate function phenomenologically. The observable consequences must then be evaluated using the complete relations specified by the model.

\end{document}